\documentclass[aps,twocolumn,showpacs]{revtex4}
\usepackage{amsmath}
\usepackage{epsfig}
\usepackage{amssymb}
\begin{document}

\title{String theory and the Taniyama-Shimura conjecture}

\newcommand*{\NJNU}{Department of Physics, Nanjing Normal University, Nanjing, Jiangsu 210097, China}\affiliation{\NJNU}

\author{Jing Zhou}\email{171001005@stu.njnu.edu.cn}\affiliation{\NJNU}
\author{Jialun Ping}\email{jlping@njnu.edu.cn}\affiliation{\NJNU}

\begin{abstract}
   The worldsheet of the string theory, which consisting of 26 free scalar fields in Minkowski space, is two dimensional
   conformal field theory. If we denote the two dimension conformal field theory by elliptic curve and denote the partition
   function of string theory by modular form, then the relation between conformal field theory and the string theory can be
   represented as the Taniyama-Shimura conjecture. Moreover, it also can be generalized to the $F$-theory.
\end{abstract}

\pacs{11.25.Hf, 11.25.Mj, 11.25.-w}

\maketitle

\section{\label{sec:introduction}Introduction}

The Langlands program of number theory, was first proposed  more or less by Robert Langlands in the late 1960s.
It is a kind of unified scheme for many results in algebra geometry and arithmetic geometry, which ranging
from quadratic reciprocity, which suggested by Gauss, to modern mathematics results such as Faltings'
proof of Mordell conjecture~\cite{GFaltings} and Wiles¡¯ proof of Fermat¡¯s last theorem~\cite{Wiles:1995ig},
which is considered as a special case of the Langlands program. However, we will not suppose any knowledge of
the number Langlands program in this work.

The Langlands duality was originally formulated in number fields, and it also can be developed analogs
in algebra curves and complex Riemann surfaces. We will focus on the geometric Langlands program
for complex Riemann surfaces here.

Interestingly, the Langlands program also appears in quantum field theory in which looks like a
very different way, known as the electro-magnetic duality. And the electro-magnetic duality can be simply
explained by Maxwell theory. For example, the classical electromagnetism can be described by the Maxwell
equations, and one may find that equations are invariant under the exchange of the electric and magnetic
fields. Then it is natural to ask whether this duality exists in the quantum field theory. In fact, it is
realized by Witten and Kaupstin through topological twisted $N=4$ super Yang-Mills
field~\cite{Kapustin:2006pk,Gukov:2006jk,Witten:2009at}.

We start with the pure $4D$ Yang-Mills theory on a Riemannian four manifold $M4$. Let $G$ be a compact
connected simple Lie group. The classical action is a functional on the space of connections on arbitrary
principal bundles on $M4$ given by the formula
\begin{equation}
 I = \frac{1}{4g^{2}}\int_{M^{4}}TrF_{A}\wedge Tr\star F_{A}+\frac{i\theta}{8\pi^{2}}\int_{M^{4}} TrF_{A}\wedge F_{A},
 \end{equation}
where $F_{A}$ is the curvature of the connection A, combining the two parameters, $g$ and $\theta$, into
one complex coupling constant, then we get
\begin{equation}
 \tau = \frac{\theta}{2\pi}+\frac{4\pi i}{g^2}.
\end{equation}

We want to focus on a ¡°topological twist¡± of this theory. There are essentially three different choices
for doing this, which is explained by Vafa and Witten~\cite{Vafa:1994tf}. The first two are similar to
Witten's construction of the topological field theory, which is topological twisted the $D=4$ $N=2$ super
Yang-Mills theory~\cite{Witten:2009at}. It is the third twist, studied in detail in Ref.~\cite{Kapustin:2006pk},
which is relevant to the geometric Langlands. For this twist, there are actually two linearly independent
operators, $Q_{l}$ and $Q_{r}$~\cite{Frenkel:2009ra}.
Then we can use any linear combination
\begin{equation}
 Q = uQ_{l}+vQ_{r}.
\end{equation}

All these are diffeomorphism-invariant. In other word, it does not depend on the Riemannian metric. To see this,
one can writes the action as,
\begin{equation}
 I = \left\{V,Q \right\}+\frac{i\Psi}{4\pi}\int_{X} TrF_{A}\wedge F_{A},
\end{equation}
where $V$ is a gauge-invariant function of the fields, and $\Psi$ is given by
\begin{equation}
 \Psi = \frac{\theta}{2\pi}+\frac{t^{2}-1}{t^{2}-1}\frac{4i\pi}{e^{2}}.
\end{equation}

As we know that the Taniyama-Shimura conjecture~\cite{LangS,Shimura_Taniyama} plays a cure role
in Wiles' proof of Fermat's conjecture~\cite{Wiles:1995ig,Wiles:19952g}, which is part of the
Langlands program. So there is a natural question to ask can we geometrize the Taniyama-Shimura conjecture?
In the following work, we show that it may shed light on this issue. In the next two sections, conformal field theory
and string theory are briefly introduced. Section IV is devote to the Taniyama-Shimura conjecture.
The summary is shown in the last section.

\section{conformal field theory}
Suppose we have a $2D$ conformal field theory~\cite{Gaberdiel:2008xb}. Then there is a representation
of left-moving and right-moving Virasoro algebras $L$ with central charges c and $\bar{c}$,
and it can write as
\begin{equation}
\left[L_{m},L_{n}\right] = \left(n-m\right) L_{m+n}+\frac{c}{12}\left(n^{3}-n\right)\delta_{n+m}.
\end{equation}
In particular, the spectrum will be assumed discrete. The most useful quantities, which we can associate to,
is the partition function. We denote $q = e^{2i\pi\tau}$, with
\begin{equation}
\tau = \theta+i\beta,
\end{equation}
and the partition function is defined as,
\begin{equation}
Z_{CFT}=\mbox{Tr}q^{L_{0}-\frac{c}{24}}\bar{q}^{\bar{L}_{0}-\frac{c}{24}}.
\end{equation}

The partition function~\cite{Billo:2006zg} has the interpretation of being the path integral on the
torus $T^{2}$ with modular parameter $\tau$. To see this we can write it as,
\begin{equation}
Z_{CFT} = e^{-2\pi\beta H+2\pi i\theta P},
\end{equation}
where the Hamiltonian $H$ is
\begin{equation}
 H = L_{0}+ \bar{L}_{0}-\frac{\left(c+\bar{c}\right)}{24},
\end{equation}
and the momentum $P$ is:
\begin{equation}
 P = L_{0}-\bar{L}_{0}-\frac{\left(c+\bar{c}\right)}{24}.
\end{equation}
The most importance of this observation is that we can now study the behavior of this theory
under diffeomorphisms of the torus $T^{2}$. As we mention above that it can be interpreted as
the path integral on a flat torus with modular parameter $\tau$, which can write as,
\begin{equation}
 E = \mathbb {C}/ \left(\mathbb {Z}+\tau\mathbb {Z}\right).
\end{equation}
Then one may note that the torus $T^{2}$ is isomorphism to the elliptic curve. In other word,
the path integral on a flat torus $T^{2}$, can be considered on the elliptic curve.

\section{String theory}
That scalar field decomposes into a zero mode and an infinite number harmonic oscillator modes.
Including the central charge, $c = 1$, the contribution from the oscillator modes is,
\begin{equation}
 \mbox{Tr} q^{L_{0}-\frac{c}{24}} = \frac{1}{q^{24}}\prod_{i=1}^{\infty}\frac{1}{1-q^{n}}.
\end{equation}
So, including both the zero mode and oscillators, we get the partition function for a
single free scalar field
\begin{equation}
 Z_{scalar}= \frac{1}{\sqrt{\left(\mbox{Im}\tau\right)}}\frac{1}{\left(q\bar{q}\right)^{24}}
 \prod_{i=1}^{\infty}\frac{1}{1-q^{n}}\prod_{i=1}^{\infty}\frac{1}{1-\bar{q}^{n}}.
\end{equation}
In lightcone gauge, there are 24 oscillator modes and 26 zero modes. Finally, we integrate over
the moduli space, then the partition function of string theory is,
\begin{equation}
\begin{aligned}
 Z_{string}= \int &\frac{1}{\left(\mbox{Im}\tau\right)}\frac{1}{\left(\alpha \mbox{Im}\tau\right)^{13}}\frac{1}{\left(q\bar{q}\right)}\left(\prod_{i=1}^{\infty}\frac{1}{1-q^{n}}\right)^{24}\\ &\left(\prod_{i=1}^{\infty}\frac{1}{1-\bar{q}^{n}}\right)^{24}\, d^{2}\tau .
\end{aligned}
\end{equation}
The function appearing in the string partition function is the Dedekind $\eta$ function~\cite{Abe:2014xja},
which it is,
\begin{equation}
\eta(q)=q^{\frac{1}{24}}\sum_{i=1}^{\infty}\left(1-q^{24}\right),
\end{equation}
It was studied long time ago by mathematicians who interested in the properties of functions
under modular transformations. one may found that the eta-function satisfies the identities,
\begin{equation}
\eta\left(\tau\right) = \eta\left(\tau+1\right),
\end{equation}
\begin{equation}
\eta \left(- \frac{1}{\tau}\right) = \eta\left(\tau\right).
\end{equation}

These two statements ensure that partition function of the scalar theory is a modular invariant
function. In terms of $\eta$, then the string partition function can be rewrite as,
\begin{equation}
Z_{string}=\int\frac{1}{\left(\mbox{Im}\tau\right)^{2}}
\left(\frac{1}{\sqrt{\mbox{Im}\tau}}\frac{1}{\eta\left(q\right)}\frac{1}{\eta\left(\bar{q}\right)}\right)^{24}d^{2}\tau .
\end{equation}
Since both the measure and the integrand are individually modular invariant, then one can prove
the string partition is also a modular function.

\section{Taniyama-Shimura conjecture}
The Shimura-Taniyama conjecture has provided a important role of much works in arithmetic
geometry over the last few decades. The Taniyama-Shimura conjecture, since known as the modularity theorem,
is an important conjecture (and now theorem) which connects topology and number theory, arising from several
problems proposed by Taniyama and Shimura. It is observed by Frey~\cite{Schimmrigk:2002upa} that the rational
solutions of Fermat curves can be used to construct certain special types of semi-stable elliptic curves.
Moreover, Jean-Pierre Serre suggested that this type Frey elliptic curve could not be modular. So, prove the
Shimura-Taniyama conjecture would therefore finally prove Fermat¡¯s last theorem. Ribet's proof~\cite{Ribet}
of Frey's conjecture, which provided key motivation for Wiles to prove the Shimura-Taniyama conjecture.
Then it indicates that the Fermat's theorem is true. More recently the work of Wiles and Taylor-Wiles
has been extended to the full Shimura-Taniyama theorem~\cite{Darmon}, removing the requirement of
semi-stability.

Let $E$ be an elliptic curve which has integer coefficients $a,b,c,d$. Let $N$ be the $j$-conductor of $E$,
let $a_{n}$ be the number appearing in the $L$-function of $E$. Then the Taniyama-Shimura conjecture says
that there exists a modular form of weight two and level $N$ whose eigenvalue under the Hecke operators
and has a Fourier series $\sum a_{n}q^{n}$.

Generally, the conjecture says that every rational elliptic curve is a modular
form~\cite{Breuil}. More formally, the conjecture suggests, for every elliptic curve,
it writes,
\begin{equation}
y^{2} = ax^{3}+bx^{2}+cx+d,
\end{equation}
over the rationals, then there exist nonconstant modular functions $f\left(z\right)$ and $g\left(z\right)$
of the same level $N$ which can be written as,
\begin{equation}
 \left[f\left(z\right)\right]^{2} = a\left[g\left(z\right)\right]^{3}+b\left[g\left(z\right)\right]^{2}+cg\left(z\right)+d.
\end{equation}
It means that every elliptic curve can be uniformized by modular function. In geometry, it is shown that
every point on the upper half plane or modular function can attach to a torus. Here one should note that
conformal field theory can be considered as elliptic curve and the partition function of string theory
$Z_{string}$ is modular invariant function, so we can rewrite Eq.$\left(21\right)$ as,
\begin{equation}
 Z_{string}^{2}=aZ_{string'}^3+bZ_{string'}^2+cZ_{string'}+d,
\end{equation}
where $Z_{string'}$ is a modular function which has same level $N$ with $Z_{string}$. Then, the relation
between conformal field theory and string theory can geometrize the Taniyama-Shimura conjecture.
Actually, in physics, conformal field theory is the worldsheet of string theory.

In fact, there is something similar arise in the $F$-theory~\cite{Vafa:1996xn,Witten:1996bn,Witten:1996qb}.
It is obtained by compactificating the type IIB string in which the complex coupling varies over the
torus~\cite{Sen:1996vd}. And the action of type IIB string is $SL\left(2,Z\right)$ invariant~\cite{Schwarz:1995dk}.
If we regard the torus as the elliptic curve, then the $F$-theory raise naturally in the Taniyama-Shimura
conjecture. The action of Type IIB has strong-weak coupling duality~\cite{Kriz:2005rf}, the modular parameter is
\begin{equation}
  \tau = i\exp^{-\phi}+C_{0}.
\end{equation}
In terms of the axio-dilaton, this is,
\begin{equation}
  \tau\rightarrow -\frac{1}{\tau}.
\end{equation}
The axionic shift symmetry is,
\begin{equation}
 C_{0} = C_{0}+1,
\end{equation}
which acts as,
\begin{equation}
 \tau\rightarrow\tau+1.
\end{equation}
Then, these generate the $SL(2,Z)$ duality group
\begin{equation}
 \tau = \frac{a\tau+b}{c\tau+d}\quad
\begin{pmatrix}
  a&b\\c&d
\end{pmatrix}
\in SL\left(2,Z\right).
\end{equation}

So the partition function of type IIB string can be considered as a modular function. By the result of
Taniyama-Shimura conjecture, we can write elliptic fibration over type IIB space-time in Weierstrass form,
\begin{equation}
 Z_{type IIB}^2 = aZ_{type IIB'}^3+bZ_{type IIB'}^2+cZ_{Type IIB'}+d,
\end{equation}
where $Z_{type IIB}$ is the partition function of type IIB string, and $Z_{type IIB'}$ is a modular
function which has the same level $N$ with $Z_{type IIB}$. In physics, it is shown that the elliptic curve
or the torus $T^{2}$ can be attached to every point of type IIB space-time, which interprets the modular
parameter as the value of axio-dilaton. And it is similar to the Sen limit~\cite{Sen:1996vd}.

\section{Summary}
In summary, we introduce the partition function of conformal field theory and string theory, which both
of them are modular function. And the conformal field theory can be considered as elliptic curve.
In fact the conformal field theory is the propagator of the string theory. It is suggested that the
relation between conformal theory and string theory can be represented as Taniyama-Shimura conjecture.
$F$-theory is obtained by compactificating the Type IIB string on the two-torous, In this case,
we suggest that it also may be described by Taniyama-Shimura conjecture. The Alday-Gaiotto-Tachikawa (AGT)
correspondence~\cite{Alday:2009aq,Alday:2009fs} hold that the partition function of 4D super Yang-Mills
is dual to the 2D conformal field theory. Then it is natural to ask the question: Can we generalize the
AGT correspondence by Taniyama-Shimura conjecture? If it is true, then it may indicate that the conjecture
is an universal principle in physics.
Actually, Taniyama-Shimura conjecture is essentially a modular theorem. The modular function is consider
as very high symmetry function. As we know, most of the partition function of string theory,
supergravity theory and super Yang-Mills theory are also satisfied some symmetry. This may be the reason
that modular theorem arises in physics.

\section*{Acknowledgment}
Part of the work was done when the first author Jing Zhou visited the Yau Mathematical Science Center.
Jing Zhou thanks for discussing with Si Li, Hai Lin and Xun Chen. This work is partly supported by the
National Science Foundation of China under Contract Nos. 11775118, 11535005.

\end{document}